\documentclass[emulateapj,natbib]{emulateapj}
\begin{document}
\title{A 5 Micron Image of $\beta$ Pictoris b at a Sub-Jupiter Projected Separation:
 Evidence for a Misalignment Between the Planet and the Inner, Warped Disk 
}
\author{Thayne Currie\altaffilmark{1}, Christian Thalmann\altaffilmark{2}, Soko Matsumura\altaffilmark{3}, 
Nikku Madhusudhan\altaffilmark{4}, Adam Burrows\altaffilmark{4}, Marc Kuchner\altaffilmark{1}}
\altaffiltext{1}{NASA-Goddard Space Flight Center}
\altaffiltext{2}{Anton Pannekoek Institute, University of Amsterdam}
\altaffiltext{3}{Department of Astronomy, University of Maryland-College Park}
\altaffiltext{4}{Department of Astrophysical Sciences, Princeton University}
\begin{abstract}
We present and analyze a new $M^\prime$ detection of the young exoplanet $\beta$ Pictoris b from 2008 VLT/NaCo data 
at a separation of $\approx$ 4 AU and a high signal-to-noise rereduction of $L^\prime$ data taken in 
 December 2009.  Based on our orbital analysis, the planet's orbit is 
viewed almost perfectly edge-on ($i$ $\sim$ 89 degrees) and 
has a Saturn-like semimajor axis of 9.50\,AU$^{+3.93\,\mathrm{AU}}_{-1.7\,\mathrm{AU}}$.
Intriguingly, the planet's orbit is aligned with the major axis of the outer disk ($\Omega$ $\sim$ 31 degrees) 
but probably misaligned with the warp/inclined disk at 80 AU often cited as a signpost for the planet's existence.  
Our results motivate new studies to clarify how $\beta$ Pic b sculpts debris disk structures and whether a second planet
is required to explain the warp/inclined disk.
\end{abstract}
\section{Introduction}
Two decades of studies have argued that the nearby, 12 Myr-old A-type star $\beta$ Pictoris likely 
harbors a young planetary system \citep[e.g.][]{SmithTerrile1984,KalasJewitt1995,Mouillet1997}; 
recently, \citet{Lagrange2009,Lagrange2010} detected a $\approx$ 9 $\pm$ 3 M$_{J}$ planet
 around this star ($\beta$ Pic b).
Imaged at projected separations of $\sim$ 6 AU and $\sim$ 8 AU (November 2003 and October--December 2009, respectively), 
$\beta$ Pic b -- along with HR 8799e \citep{Marois2011,Currie2011} -- may also provide a more direct comparison 
to the solar system's gas giants than other directly imaged planets which are 
at wider separations \citep[e.g. Fomalhaut b and HR 8799bcd][]{Kalas2008,Marois2008}.  
Current studies have yet to detect the planet at projected separations $\lesssim$ 0.3$\arcsec{}$ \citep{Lagrange2009b,Fitzgerald2009}.  
Data at these smaller separations could provide crucial constraints on the planet's orbit.

Imaging planets at small, $<$ 0.3$\arcsec{}$ separations requires significantly reducing quasi-static speckle noise 
and wavefront errors induced by imperfect AO corrections.  Advanced observing/image 
processing techniques like \textit{angular differential imaging} (ADI) coupled with PSF subtraction 
from a \textit{locally optimized combination of images} (LOCI) algorithm \citep{Marois2006,Lafreniere2007a} 
significantly attenuate speckles and increase sensitivity.  New instrumentation, such as the 
\textit{Gemini Planet Imager} \citep[GPI][]{MacIntosh2008}, will achieve far superior wavefront 
control.  Generally, these efforts focus on planet imaging in the near-IR.
However, mid-IR imaging naturally overcomes some of these challenges as the achievable Strehl ratio 
is better and the planet-to-star contrast is most favorable.  
High-contrast imaging with Strehl ratios $\ge$ 0.9 can yield at least some planet detections close 
to the telescope diffraction limit \citep[e.g. 2 $\lambda$/D for HR 8799d;][]{Serabyn2010}.
Because $M$ band imaging often achieves these high Strehl ratios \citep{Minowa2010,Hinz2006}, it may be
a promising route for detecting very young and self luminous planets at small $\lambda$/D separations, 
despite a much higher sky background.

In this Letter, we report a detection of $\beta$ Pic b at a separation of $\sim$ 0.21$\arcsec{}$ extracted 
from archival $M^\prime$ band VLT/NaCo data taken in November 2008.
We also present a high signal-to-noise $L^\prime$ detection of $\beta$ Pic b from December 2009 data first published 
in \citet{Lagrange2010}.
We combine these data with recent data from \citet{Bonnefoy2011} and \citet{Quanz2010} to 
better constrain the orbit and atmosphere of $\beta$ Pic b. 

\section{Observations and Data}
Our study originates from the need to test the ADI/LOCI reduction pipeline first presented in \citet{Currie2010a} and 
updated in \citet{Currie2011} at separations smaller than those where the pipeline had previously extracted 
planet signals (r $<$ 0.375$\arcsec{}$).  
%The current version adds features useful for detecting and characterizing planets 
%in the mid-IR and at small separations: a more robust sky subtraction method and a robust measure of the astrometric 
%bias in separation and position angle induced by point source self subtraction.
%The major updates to the current code used on 
%VLT/NaCo data for HR 8799 in \citet{Currie2011} include the following: a robust pixel-by-pixel outlier rejection for 
%data taken in cube mode; a more robust means to identify and interpolate over remaining hot/cold pixels using a 
%moving box median filter; a robust measure of the astrometric bias induced by point source self subtraction; 
%and better flexibility in removing the smooth seeing halo, selecting annuli for 
%PSF subtraction, and masking/unmasking the diffraction spider and latent features.  
%, the sensitivity gain of a LOCI-focused reduction 
%pipeline compared to simple PSF subtraction is strongest at small inner working angles \citep{Lafreniere2007a}.  
Because $\beta$ Pic b's reported projected separation in 2009 was $\approx$ 0.3$\arcsec{}$ \citep{Lagrange2010}, we chose the now publicly 
available \citeauthor{Lagrange2010} $L^\prime$ band data from December 29, 2009 to test our code performance.
\citet{Lagrange2010} discusses the details of the $L^\prime$ band observations.
The total field rotation in units of the image FWHM was $\sim$ 3 $\lambda$/D, sufficient 
for using our reduction pipeline.
% for $L^\prime$ band and xxx $\lambda$/D for
%Ks band.

\begin{figure}
\centering
%\epsscale{0.5}
%\plotone{betapic_lpv3.eps}
%\\
\plottwo{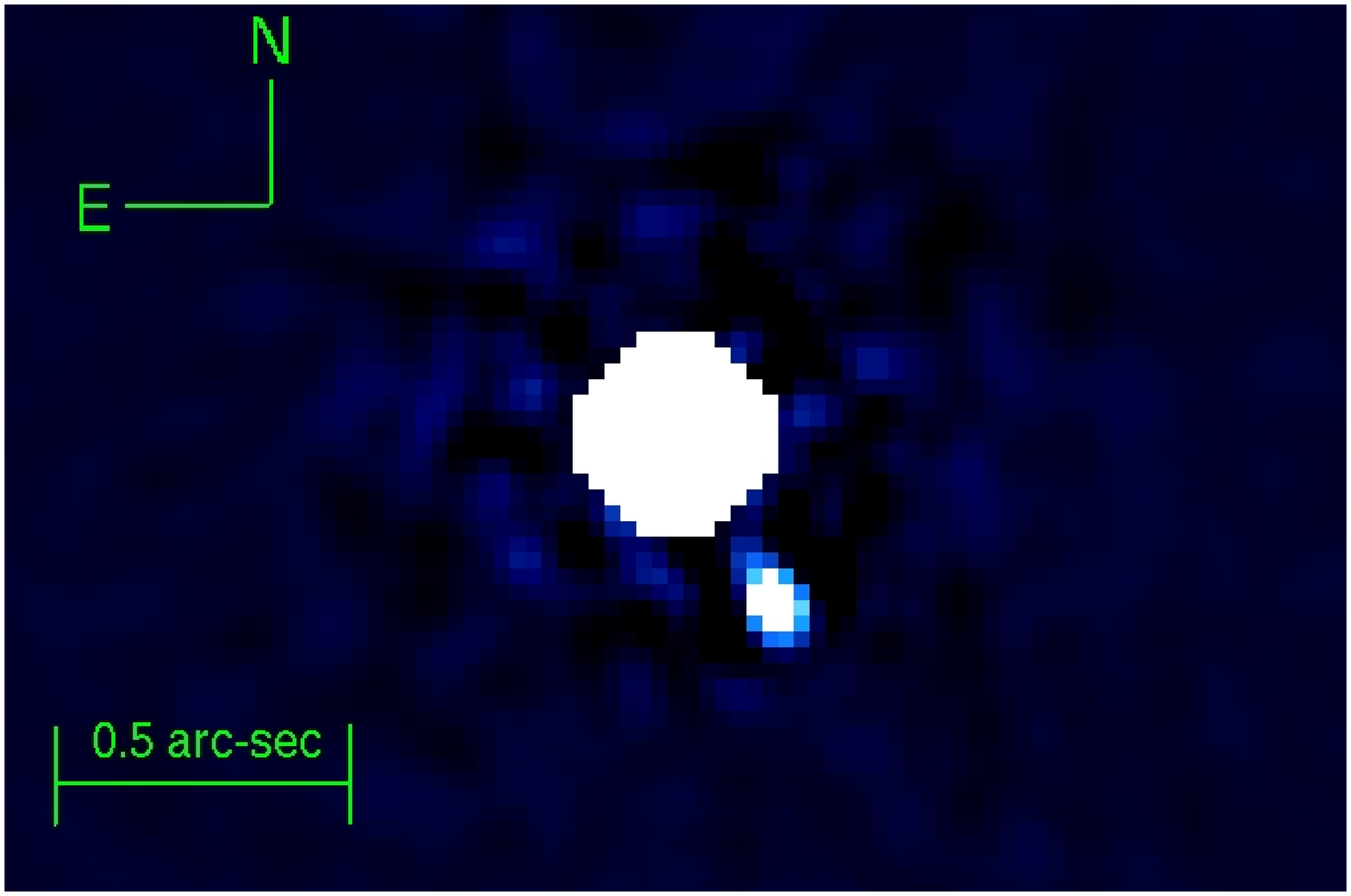}{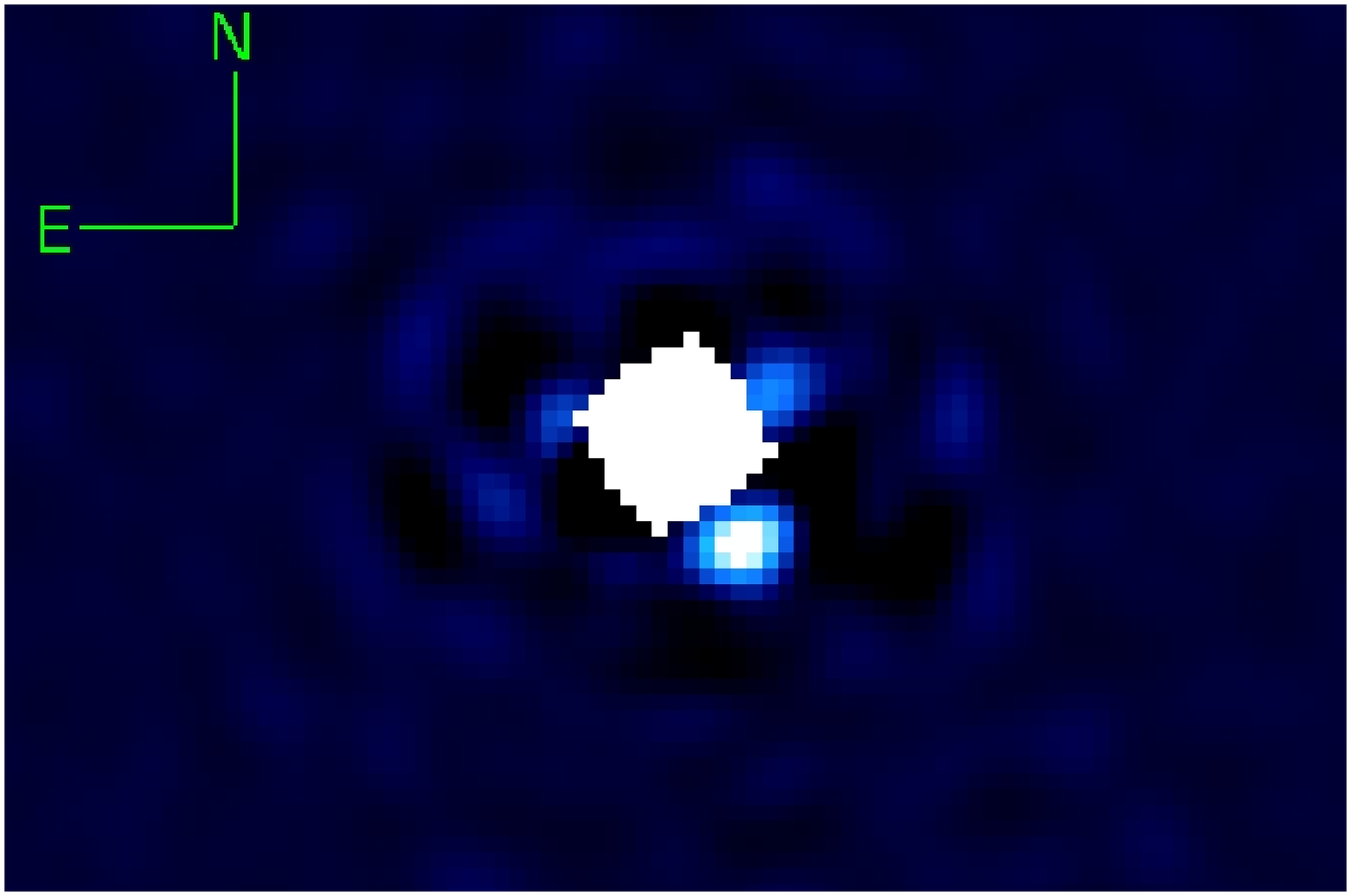}
\epsscale{1}
%\plottwo{betapic_mpv2.eps}{betapic_mpv2snr.eps}
\plottwo{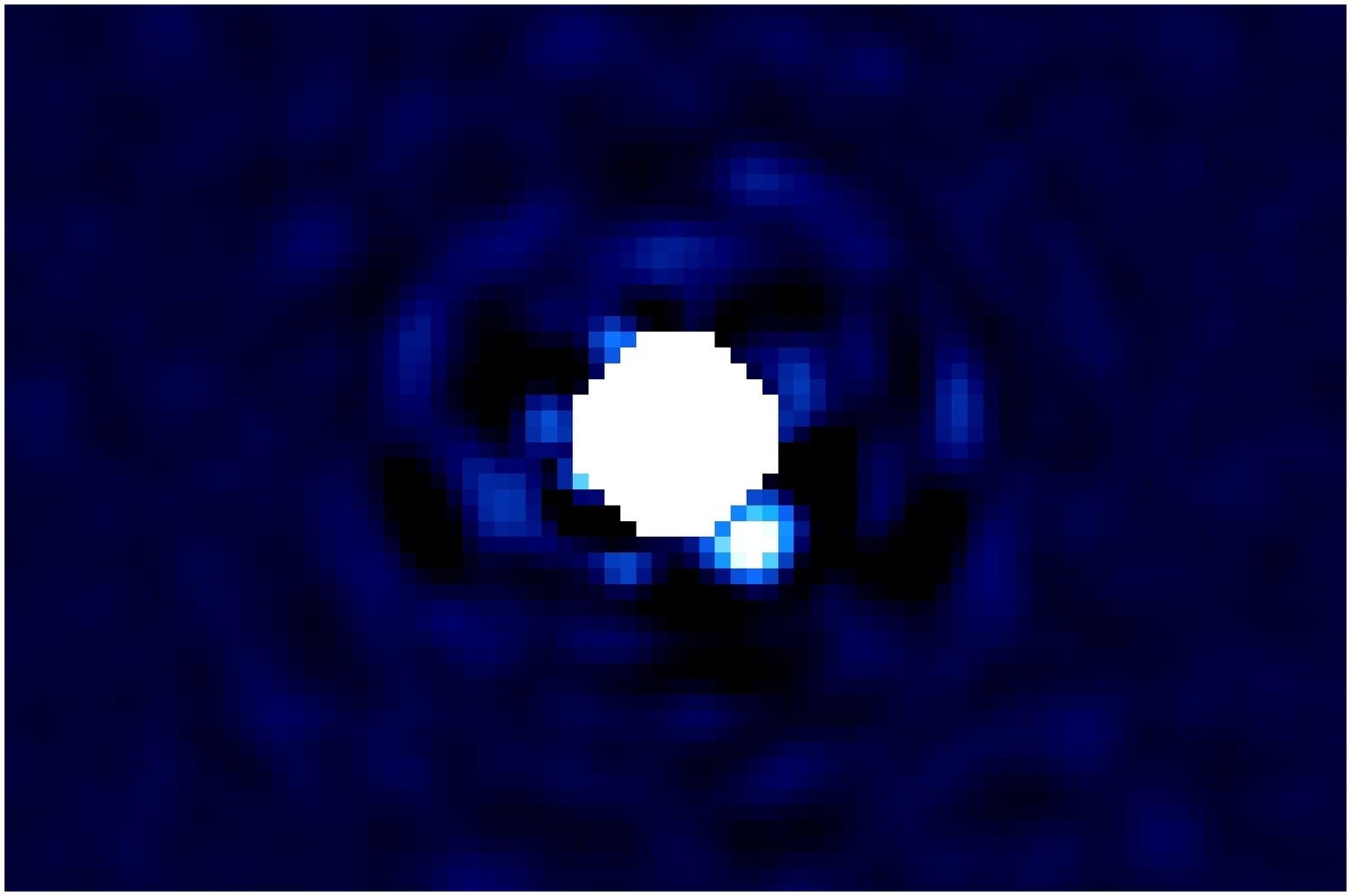}{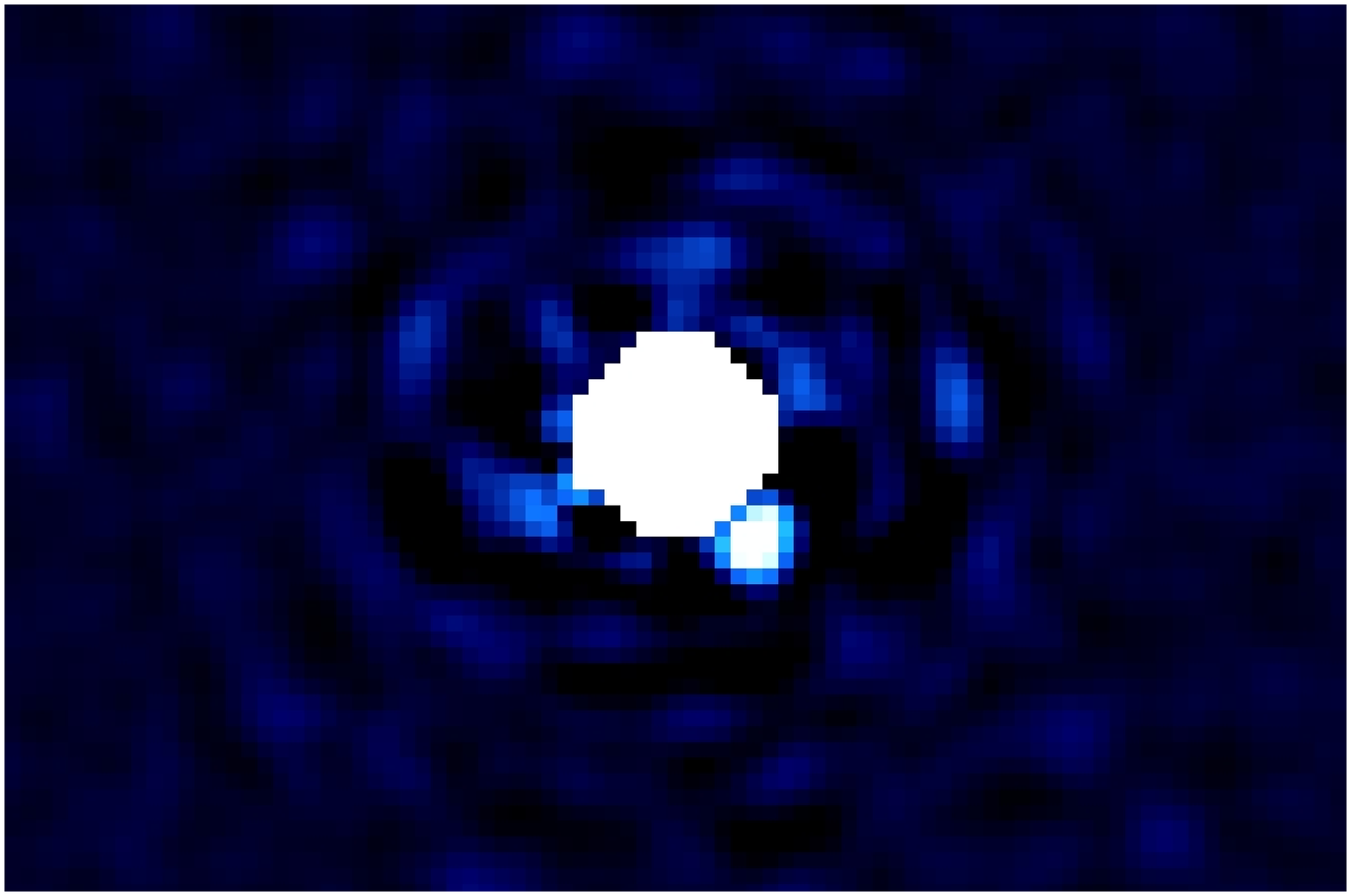}
\caption{(Top-left) Reduced VLT/NaCo $L^\prime$ band image from Dec. 29, 2009 data showing a 21-$\sigma$ detection of $\beta$ Pic b 
at a separation of $\sim$ 0.32".  
(Top-right) Reduced image showing the detection of $\beta$ Pic b in $M^\prime$ band (same image size).
The LOCI parameters used to construct the reduced image include $\delta$ = 0.25, $g$ = 0.4, and $N_{A}$ = 250$\times$FWHM.  
(Bottom panels) Reduced images using less aggressive LOCI settings -- $\delta$ = 0.5, $g$=0.5, and $N_{A}$ = 300$\times$FWHM 
(left panel) and $\delta$ = 0.5, $g$=1, and $N_{A}$ = 300$\times$FWHM (right panel) -- yielding detections of 
SNR $\sim$ 5 and 4.5, respectively.  
In general, we detect the $\beta$ Pic planet at a 4--6 $\sigma$ level using a range of LOCI parameters:
 $\delta$ = 0.25--0.5, $g$ = 0.3--1, $N_{A}$ = 200--300.}
\label{bpicimage}
\end{figure}

Figure \ref{bpicimage} (top-left panel) shows our processed $L^\prime$ band image 
 using the LOCI algorithm in annular regions of 250$\times$FWHM ($N_{A}$ = 250) with reference images selected 
from frames with at least 0.5$\times$FWHM field rotation ($\delta$ = 0.5).  The planet is 
easily detected and is well separated from residual speckle noise.
The planet signal-to-noise, determined from the dispersion in pixel intensity 
values in concentric annuli, is SNR $\sim$ 21: about a factor of 4--5 greater than from \citet{Lagrange2010}.  

Motivated by this success, we searched for additional $\beta$ Pic data in the VLT/NaCo archive taken in ADI mode
 between 2003 and 2009, finding a set taken on November 11, 2008 with the L27 camera.  Most of these data were taken in 
sparse aperture masking mode in $K_{s}$, $L^\prime$, and $M^\prime$ bands over a span of $\sim$ 4 hours: 
these data were mentioned in \citet{Lagrange2009b} as not providing good constraints on the companion.  However, we 
found $\sim$ 13 minutes of the $M^\prime$ data taken in ADI mode without aperture masking at various times 
in between the masking data.  Over the course of the entire observing sequence, the 
parallactic angle changed by $\sim$ 100 degrees, or $\sim$ 2.4--3 $\lambda$/D at 0.2$\arcsec{}$--0.25$\arcsec{}$: 
 sufficient for image processing with our pipeline.   

Basic image processing of the $M^\prime$ band data followed steps outlined in \citet{Currie2011}.
%, including pixel-by-pixel 
%outlier rejection, sky subtraction, bad pixel identification, and subpixel registration. 
After registering each image and subtracting off the smooth seeing halo, we Fourier filtered 
the data to remove residual low spatial frequency noise and masked
any hitherto unidentified bad pixels previously lost in the seeing halo.  We explored a range of LOCI parameter space, 
varying $\delta$, $N_{A}$, and the ratio of the radial to azimuthal lengths of the subtraction annulus ($g$).
  Because $\beta$ Pic b is very luminous in the mid-IR \citep[e.g. $\Delta$$L^\prime$ $\approx$ 7.7;][ 
and Section 3 of this work]{Lagrange2010}, we focused on ``aggressive" LOCI settings of $\delta$ = 0.25--0.5, $N_{A}$ = 200-300, 
and $g$ = 0.3--1, which better remove residual speckle noise.

Figure \ref{bpicimage} (top-right panel) shows our best reduced $M^\prime$ image.  
  $\beta$ Pic b is clearly detected in the southwest quadrant
  $\approx$ 0.2--0.25$\arcsec{}$ from the star (SNR $\sim$ 6).  
Manually inspecting each image between the radial profile subtraction 
and final image combination steps and examining a signal-to-noise map of the median-combined image 
also shows that the peak does not result from latent image artifacts.
Slightly different settings for $\delta$, $N_{A}$, and $g$ also yield significant detections (bottom panels).

\section{Analysis}
%\subsection{Astrometric Calibration}
%\section{Astrometric Analysis and Orbit Fitting}
Our new $M^\prime$ band detection and high signal-to-noise $L^\prime$ band detection 
allow new constraints on the planet's orbit.  To derive precise astrometry needed to investigate the planet's 
orbit, we adopt the NaCo plate scale and orientation for the L27 camera from \citet{Bergfors2011}: 27.1 mas/pixel and a 
north position angle of -0.6 degrees.  These values are nearly identical to those for the L27 camera quoted by \citet{Lagrange2009} for 2003 NaCo 
data and for the S27 camera from \citet{Ehrenreich2010} calibrated from Trapezium data acquired closest in time to the $\beta$ Pic data: 
our astrometric results does not leverage on which calibration we use.

To fine tune our measurements, 
%we consider the photometric and astrometric bias induced 
%by LOCI processing due to partial point source self subtraction.  
%While this bias is unimportant for planets imaged at wide separations with observations covering signficant field rotation, it may be 
%important for $\beta$ Pic b.  
%The flux loss from self subtraction is severe at small separations like $\beta$ Pic b's \citep{Lafreniere2007a}. Furthermore,
% the \textit{rate of change of the flux loss} is also larger \citep[see figures in ][]{Lafreniere2007a,Currie2011}, which may bias
 %the radial component of the computed centroid position.
we correct for the photometric and astrometric bias induced by LOCI processing by comparing the imputed fluxes and positions of 
fake point sources added to registered images with computed fluxes and centroid positions obtained after LOCI processing
 \citep[e.g.][]{Lafreniere2007a,Thalmann2009,Currie2011}.
While we lack unsaturated data from this run to directly confirm the PSF shape, unsaturated $M^\prime$ data taken in prior runs 
such as that for HD 158882 (March 2007) show that the AO-assisted NaCo $M^\prime$ PSF core is axisymmetric and well reproduced by a simple Gaussian 
intensity distribution.  For the $L^\prime$ band data, the astrometric bias is minimal, whereas $\beta$ Pic b's measured 
radial separation in $M^\prime$ band is biased by about +0.5 pixels (0.013").
The position angle offsets for both data were minimal.

%After correcting for these biases, 
We determine the $M^\prime$ band 
position to be at a separation of r = 0.210 $\pm$ 0.027$\arcsec{}$ and position angle of  
211.49 $\pm$ 1.9 degrees.  The $L^\prime$ band position is at 0.326 $\pm$ 0.013$\arcsec{}$ and 210.64 $\pm$ 1.2 degrees (Table 1).  
Here we conservatively 
assume an uncertainty in radial separation of one pixel for $M^\prime$ band and 0.5 pixels for the (higher signal-to-noise) $L^\prime$ band data.  The position angle 
uncertainty -- determined from the dispersion in values using different centroiding estimates (e.g. \textrm{cntrd.pro} vs. \textrm{gcntrd.pro}) is 0.25 pixels 
(0.7 mas$\times$r), or 1.2 and 1.9 degrees for $L^\prime$ and $M^\prime$, comparable to uncertainties for 
$\beta$ Pic b by \citet{Lagrange2009,Lagrange2010} and \citet{Bonnefoy2011}.  Assuming that $\beta$ Pic is 19.3 pc distant
\citep{Crifo1997}, the planet was at a 
projected separation of 4.05 $\pm$ 0.50 AU on November 11, 2008 and 
6.29 $\pm$ 0.25 AU on December 29, 2009.
%\subsection{Orbit Fitting}
%\section{Constraints on the Orbit of the $\beta$ Pic Planet}

To determine the range of allowable orbits for $\beta$ Pic b, 
%two sets of analysis.  First, we investigate a simple case, assuming a circular orbit 
%and determining the range of other orbital parameters -- semimajor axis (a), inclination (i), 
%longitude of ascending node ($\Omega$), etc. -- from the set of configurations 
%with a $\chi^{2}_{\nu}$ $<$ 14 (chi$^{2}$ $\le$ 3 $\sigma$).  The best-fitting model (Figure xxx) 
%has a$_{p}$ = 10.5 AU, i = 89.1 degrees, and $\Omega$ = xxx.  The range of models 
%with $\chi^{2}$ $<$ 14 cover a$_{p}$ = 8--20 AU, i = 85--90 degrees, and $\Omega$ = xxx-xxx.  
%Second, following \citet{Thalmann2009} and \citet{Currie2010}, we perform a Monte Carlo simulation comparing 
we follow the method described in \citet{Janson2011} used to model the orbit of
the low-mass brown dwarf companion GJ 758 B \citep{Thalmann2009,Currie2010a}, somewhat 
similar to earlier analyses for $\beta$ Pic b \citep{Lagrange2009b,Fitzgerald2009}.
In this approach, we perform a Monte Carlo simulation comparing
the astrometry to predictions from randomly selected orbits, where we allow all orbital parameters 
to vary.  The minimum $\chi^{2}$ value in our simulation is $\chi^{2}$ $\sim$ 1.23.  
Given our data's weak constraints on the orbital acceleration and the
degeneracies due to the unknown line-of-sight components of planet
position and velocity, no single 'best' orbital solution emerges.
Rather, the best-fitting solutions describe an extended, well-
defined family of solutions that all match the data equally well.
We choose a cut of $\chi^{2}$ $\le$ 2.23 ($\chi^{2}$ $\lesssim$ $\chi^{2}$$_\mathrm{min}$
 + 1) to represent the family of best-fitting orbits.  We also consider the results
for a cut of $\chi^{2}$ $\le$ 8 ($\chi^{2}$ for a 1-$\sigma$ deviation in each of the
data's degrees of freedom): a family of 'average- fitting' solutions.
From the set of models satisifying this criterion, we determine the 
median value of each parameter, weighted by the ratio of the mean to current orbital velocity 
for the corresponding orbit, and identify the weighted 68\% confidence interval about the median.
We include astrometry from the highest signal-to-noise data separated in time by more than 
$\sim$ 3 months (Table 1).

\begin{figure}
\centering
\epsscale{0.85}
\plotone{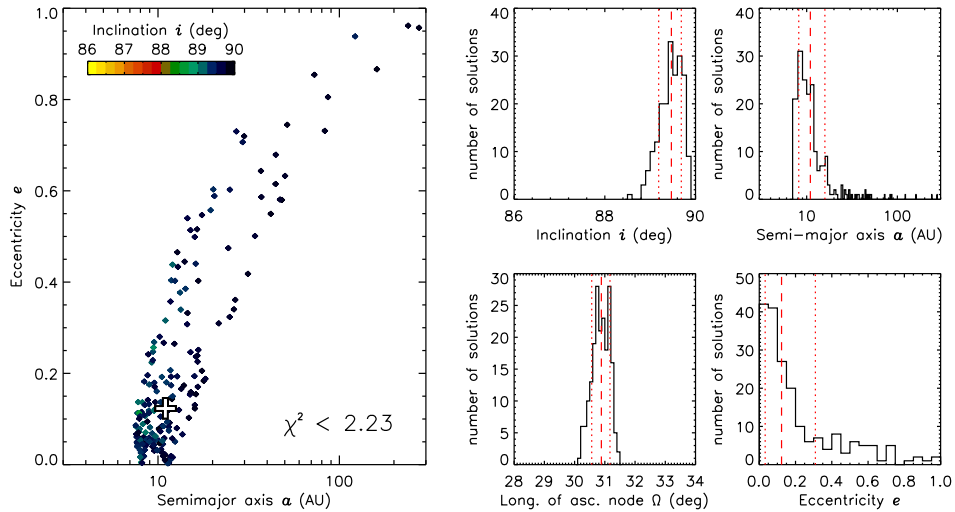}
%\plotone{orb_cut8_panels.eps}
\caption{Orbital analysis results following the Monte Carlo method outlined in \citet{Thalmann2009}, 
showing parameters for solutions fulfilling $\chi^{2}$ $\le$ 2.23.  The 
left panel displays the family of solutions in semimajor axis, eccentricity, and inclination space.  
The right panels show histogram distributions of these three parameters and the longitude of the 
ascending node.  The vertical dashed line identifies the weighted median value for each parameter; the 
vertical dotted lines define the 68\% confidence interval for each parameter.}
\label{orbitfit}
\end{figure}

\begin{figure}
\centering
\plotone{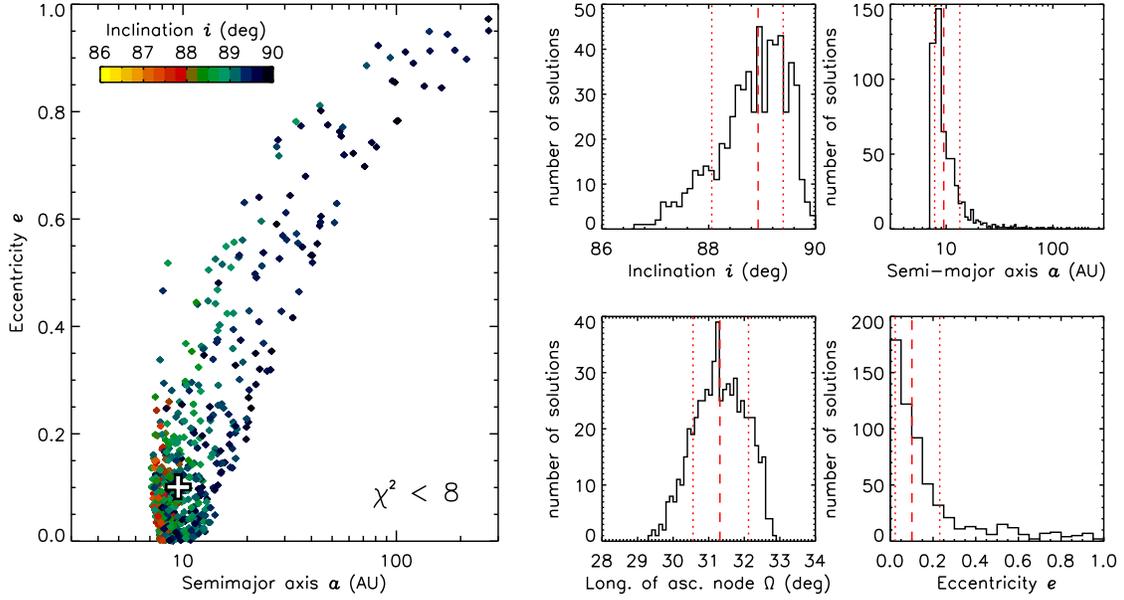}
\caption{Same as previous figure except for orbital solutions fulfilling $\chi^{2}$ $\le$ 8.}
\label{orbitfit2}
\end{figure}

Figures \ref{orbitfit} and \ref{orbitfit2} displays our 
Monte Carlo simulation results.  For a $\chi^{2}$ $\le$ 2.23 
cutoff (Figure \ref{orbitfit}), the range of 
best-fit orbital parameters (weighted median, [weighted 68\% confidence interval]) 
include $a_{p}$ = 10.99 [8.18, 15.88] AU, $i$ = 89.47 [89.19, 89.69] degrees, 
$e$ = 0.12 [0.03,0.31], and $\Omega$ = 30.89 [30.57, 31.17] degrees.  
For nearly circular orbits ($e$ $<$ 0.1), 
the range in semimajor axes is much narrower ($a_{p}$ $\sim$ 8--12 AU).  If we relax our 
fitting criteria to accept models with $\chi^{2}$ $\le$ 8 (Figure \ref{orbitfit2}), 
we find $a_{p}$ = 9.50 [7.80, 13.43 ] AU and $e$ = 0.10 [0.02,0.23].
More importantly, the inclination and longitude of ascending node are still 
nearly single valued: $i$ = 88.93 [88.06, 89.40] and $\Omega$ = 31.32 [30.56, 32.12].
Thus, $\beta$ Pic b's orbit must be viewed almost perfectly edge on, consistent with that for 
$\beta$ Pic's debris disks, with a north position angle of $\sim$ 30.8 degrees for the outer debris disk \citep[see also 
Boccaletti et al. 2009]{KalasJewitt1995} but 
 inconsistent with the inner disk position angle, which is offset by $\sim$ 5 degrees \citep{Heap2000,Golimowski2006}.

%\section{Atmospheric Modeling}
Unfortunately, the $M^\prime$ band observations of $\beta$ Pic were taken with the star saturated within $\sim$ 3 pixels ($\sim$ 0.6 FWHM) 
of the centroid and there were likewise no unsaturated standard star observations.  We derive a very crude magnitude estimate by 
scaling the $M^\prime$ PSF of HD 158882 to the unsaturated portion of the $\beta$ Pic PSF and use HD 158882's known brightness 
($K_{s}$ = 5.09; $K_{s}$-$M^\prime$ $\sim$ 0) to calibrate $\beta$ Pic b's brightness.  We estimate 
$\Delta M^\prime$ $\approx$ 8.02 $\pm$ 0.50 (M$_{M^\prime}$ $\approx$ 9.99), where we consider the uncertainties in our PSF fitting 
scaling, the dispersion in individual planet magnitude estimates drawn from separate reductions, and 
the intrinsic signal-to-noise of our detection.  We determine an $L^\prime$ contrast of $\Delta$$L^\prime$ = 7.71 $\pm$ 0.06.
  Combining the $L^\prime$ measurement with the $K_{s}$ band and [4.05] data from \citet{Bonnefoy2011} and \citet{Quanz2010}, we have three 
good quality photometric points to investigate the family of possible solutions for $\beta$ Pic b's atmospheric properties.  

\begin{figure}
\centering
\epsscale{0.65}
\plotone{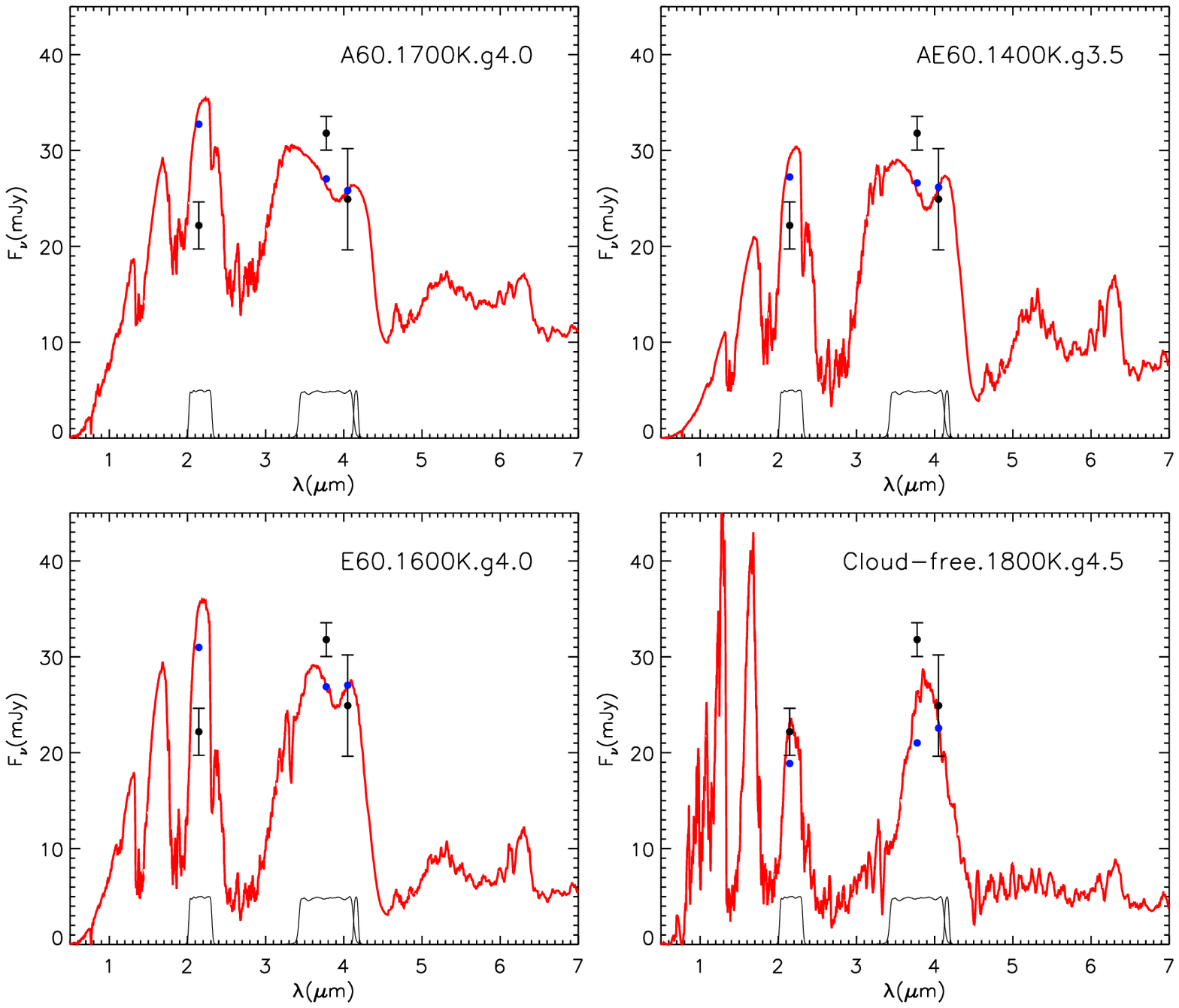}
\caption{Figures comparing the $\beta$ Pic b photometry to spectra for a range of cloud prescriptions: 
the Model A 'thick cloud' limit used in \citet{Currie2011} (upper-left), the Model AE 'thick cloud' 
prescription from \citet{Madhusudhan2011} (upper-right), the Model E brown dwarf cloud prescription 
from \citet{Burrows2006} (lower-left), and a cloud-free atmosphere also from \citet{Burrows2006} (lower-right). 
The blue dots show the flux predictions for each bandpass.
All models assume a modal particle size of 60 $\mu m$.  
%None of the models provide exemplary fits to the data, indicating 
%that the planet has parameters (log(g), temperature, cloud prescription) lying outside the range we 
%considered.
}
\label{atmosfit}
\end{figure}

Figure \ref{atmosfit} compares the $\beta$ Pic b photometry to best-fit spectra for models with log($g$) = 3.5/4--4.5 and $T_{eff}$ = 
1000--1800 K for a range of cloud prescriptions: the Model A and AE thick cloud prescriptions 
respectively from \citet{Currie2011} and \citet{Madhusudhan2011} that best fit
 the HR 8799 planet SEDs, the Model E cloud deck prescription appropriate for brown dwarfs \citep{Burrows2006}, and a 
cloudless atmosphere.  The $\chi^{2}_{\nu}$ values for these models are, respectively,
$\chi^{2}_{\nu}$ = 24.8, 12.3, 20.4, and 43.2 for Models A, AE, E, and the cloudless case.  
The AE thick cloud model provides the best fit.
Thick cloud models also produce redder $L^\prime$-$M^\prime$ colors at high temperatures, 
similar to that estimated here ($\approx$ -0.22 $\pm$ 0.50), though our lack of a reliable 
$M^\prime$ photometric calibration precludes strong conclusions.
  Good photometry is available in only three filters, so we cannot yet say that
 $\beta$ Pic b has thick clouds like the HR 8799 planets \citep{Currie2011, Madhusudhan2011}. 

The range of gravities and effective temperatures are log(g) = 3.5--4.5 and $T_{eff}$ = 1400--1800 K.  
The implied masses for these models range between 4.1 M$_{J}$ and 19.2 $M_{J}$
and ages range between 1 and 27 Myr, broadly consistent with the planet mass \citep[9 $\pm$ 3 M$_{J}$;][]{Lagrange2010},
  stellar age \citep[12 Myr;][]{Zuckerman2001}, and likely formation timescale \citep[$\le$ 3--5 Myr;][]{Currie2009}.
Planet fluxes in the near-IR (1--1.65 $\mu$m) and the 3--3.5 $\mu m$ methane absorption trough
are highly sensitive to cloud structure \citep[e.g.][]{Currie2011,Madhusudhan2011}. Thus, $J$ or $H$ broadband data and/or narrowband 
3--3.5 $\mu m$ data will be critical in breaking model fitting degeneracies.
 
\section{Discussion}
We present a new detection of $\beta$ Pic b in $M^\prime$ band at a separation of
$\sim$ 0.21$\arcsec{}$ (a$_\mathrm{projected}$ = 4.05 AU) from archival VLT/NaCo data taken in November 2008 and 
a high SNR rereduction of $L^\prime$ data first reported by \citet{Lagrange2010}, using 
these data to constrain the planet's orbit and atmospheric properties.  For orbits whose fit to the 
data yield $\chi^{2}$ $\le$ 2.23, we find that the $\beta$ Pic planet has 
a semimajor axis of $a_{p}$ = 10.99 AU$^{+4.69 \mathrm{AU}}_{-2.81 \mathrm{AU}}$ 
and a moderate/low eccentricity ($e$ $\lesssim$ 0.31).  
Admitting orbital solutions with $\chi^{2}$ $\le$ 8, the parameter ranges are 
$a_{p}$ = 9.50 AU$^{+3.93 \mathrm{AU}}_{-1.7 \mathrm{AU}}$ and e $\lesssim$ 0.23.  
In both cases, values for the planet's inclination (i $\sim$ 88.06--89.69 degrees) 
and longitude of ascending node ($\Omega$ $\sim$ 30.56--32.12 degrees) 
are tightly constrained and imply that the planet's orbit is almost perfectly 
aligned with the outer debris disk, but not the inclined inner disk ($\Omega$ $\sim$ 35--36 
degrees).  We cannot extract reliable photometry from our $M^\prime$ band data; new data 
at 1--1.65 $\mu m$ and $\sim$ 3--3.5 $\mu m$ is needed to constrain $\beta$ Pic b's atmosphere.

Numerous studies of the $\beta$ Pic debris disk(s) have identified the star as harboring a 
young planetary system \citep[e.g.][]{SmithTerrile1984,KalasJewitt1995,Mouillet1997,Weinberger2003}.  More recently, 
the presence of a warp in the disk at $\sim$ 80 AU -- due to the combined effects of the main disk 
with PA $\sim$ 30.8 and a second, inclined disk offset by 5 degrees -- was identified as a clear 
signpost of a perturbing planet \citep[e.g.][]{Mouillet1997,Heap2000,Golimowski2006}, motivating high-contrast imaging studies 
to image the planet.  \citet{Lagrange2009} identified $\beta$ Pic b as a likely source of the inclined disk and used the disk morphology
to derive mass estimates \citep[see also][]{Lagrange2010}.

Our results suggest that $\beta$ Pic b is probably \textit{not} aligned with the inner disk/warp but rather the main disk, as the 
allowed range in $\Omega$ is offset from the main disk as measured by \citet{KalasJewitt1995} by no more than $\sim$ 1 degree. 
Furthermore, the planet may be misaligned with the submm disk emission \citep{Wilner2011}, which 
is sensitive to dynamical sculpting by planets \citep{Kuchner2010}.
However, models accounting for the inclined inner disk presume that the planet's orbit 
is also inclined relative to the main disk \citep[e.g.][]{Mouillet1997,Augereau2001}. 
New beta Pic b astrometry, a more precise astrometric 
calibration of existing beta Pic NaCo data by determining and correcting for image 
distortion, and a detailed relative calibration between NaCo data and data revealing 
the disk will further clarify how beta Pic b's orbital plane compares to that for the main disk and the inner disk/warp.
Furthermore, our new orbital constraints for $\beta$ Pic b strongly motivate new studies of the 
dynamical sculpting of $\beta$ Pic's debris disk by planets.   If $\beta$ Pic b 
or non-planet related mechanisms \citep[e.g.][]{Armitage1997} fail to explain the inclined 
debris disk/warp, the existence of additional planets in the system may be required. 

Our $M^\prime$ band detection demonstrates that it is possible to directly image planets at separations approaching 
the telescope diffraction limit without sparse aperture masking interferometry \citep[SAM;][]{Ireland2008}.  
The high Strehl ratio, large amount of field rotation, large mid-IR planet brightness, and 
LOCI processing pipeline are the keys to closing this 
gap.  While SAM can detect planets interior to the telescope diffraction limit, it is overall less sensitive.
%, having 
%failed to detect some planets known through direct imaging \citep[e.g. HR 8799bcde;][]{Hinkley2011}.
However, the techniques can be complementary, yielding
detections or robust limits on infant gas giant planets around the youngest stars on $\sim$ 5--100 AU scales.

Upcoming facilities like GPI, SPHERE, SCExAO, and Project 1640 achieve higher contrast at small inner working angles 
in the near-IR primarily through more sophisticated wavefront control \citep{MacIntosh2008,Beuzit2008,Martinache2009,Hinkley2011}.  
Our results, coupled with previous $L^\prime$ band detections of $\beta$ Pic from \citet{Lagrange2009,Lagrange2010} 
and the high signal-to-noise $L^\prime$ band detection of HR 8799e \citep{Marois2011} suggest that the mid-IR may 
also be fertile ground for new exoplanet detections at small separations for very young 
systems.  Young, nearby 1.5--2 $M_{\odot}$ stars like $\beta$ Pic are particularly promising targets for direct imaging 
surveys \citep[e.g.][]{Crepp2011} and many have resolved debris disks \citep[e.g. HD 181327,][]{Schneider2006, Chen2008}.  
Imaging massive planets in such systems can yield additional studies of planet-disk interactions, such as those 
motivated by this work.  

\acknowledgements We thank David Ehrenreich, Karl Stapelfeldt, Scott Kenyon, Justin Crepp, and the anonymous referee for suggested 
improvements to the manuscript and Michael McElwain, Sally Heap, Sarah Maddison, and Aki Roberge for other useful discussions.
TC is supported by a NASA Postdoctoral Fellowship.
AB is supported in part under NASA ATP grant 
NNX07AG80G, HST grant HST-GO-12181.04-A, and JPL/Spitzer Agreements 
1417122, 1348668, and 1371432.  SM is supported by an Astronomy Center for Theory and Computation
Prize Fellowship.  

{}

\begin{deluxetable}{llllllllllllll}
\tablecolumns{2}
\tablecaption{$\beta$ Pic Data Used in this Paper}
\scriptsize
\tablehead{}
\startdata
\textbf{Astrometry}\\
Date&Filter &Separation ($\arcsec{}$),Position Angle ($^{\circ}$)&Reference\\
\hline
11-10-2003& $L^\prime$&0.411 $\pm$ 0.008, 31.7 $\pm$ 1.3 & \citet{Lagrange2009}\\
11-11-2008& $M^\prime$&0.210 $\pm$ 0.027, 211.49 $\pm$ 1.9 & this work\\
12-29-2009& $L^\prime$&0.326 $\pm$ 0.013, 210.64 $\pm$ 1.2 & this work\\ 
03-20-2010& $K_{s}$&0.345 $\pm$ 0.012, 209.8 $\pm$ 0.8 & \citet{Bonnefoy2011}\\
\textbf{Photometry}\\
Date&Filter &Absolute Magnitude&Reference\\
\hline
03-20-2010&$K_{s}$& 11.20 $\pm$ 0.12 & \citet{Bonnefoy2011}\\
12-29-2009&$L^\prime$ & 9.73 $\pm$ 0.06 & this work\\
04-03-2010&$[4.05]$&9.77 $\pm$ 0.23 & \citet{Quanz2010}\\
11-11-2008&$M^\prime$&$\approx$ 9.99 $\pm$ 0.50 & this work
%\textbf{Derived Orbital Properties}\\
%Semimajor Axis (AU) & 44.12 (30.28, 76.66) \\
%Eccentricity, Inclination & 0.73 (0.52, 0.85), 49.64$^{o}$ (27.87$^{o}$,63.34$^{o}$)\\
\enddata
\tablecomments{ Our astrometry and photometry are drawn from the three separate reductions shown 
in Figure 1.
The $L^\prime$ measurement assumes L$^\prime_{\beta,\ Pic}$ = 3.45.  
The $M^\prime$ photometry lacks reliable photometric calibration and thus is not useful for atmospheric modeling.  
}
\end{deluxetable}



\begin{thebibliography}{}
%\bibitem[Allard et al.(2001)]{Allard2001}Allard, F., et al., 2001, \apj, 556, 357
%\bibitem[Baraffe et al.(2003)]{Baraffe2003}Baraffe, I., et al., 2003, A\&A, 402, 701
\bibitem[Armitage and Pringle(1997)]{Armitage1997}Armitage, P., Pringle, J., 1997, \apj, 488, 47L
\bibitem[Augereau et al.(2001)]{Augereau2001}Augereau, J. C., 2001, A\&A, 370, 440
%\bibitem[Barman et al.(2011)]{Barman2011}Barman, T., et al., 2011, \apj\ in press
\bibitem[Bergfors et al.(2011)]{Bergfors2011}Bergfors, C., et al., 2011, A\&A, 528, 134
\bibitem[Beuzit et al.(2008)]{Beuzit2008}Beuzit, J.-L., et al., 2008, SPIE, 7014, 41
%\bibitem[Biller et al.(2007)]{Biller2007}Biller, B., et al., 2007, \apjs, 173, 143
%\bibitem[Biller et al.(2010)]{Biller2010}Biller, B., et al., 2010, \apj, 720, 82L
\bibitem[Boccaletti et al.(2009)]{Boccaletti2009}Boccaletti, A., et al., 2009, A\&A, 495, 523 
\bibitem[Bonnefoy et al.(2011)]{Bonnefoy2011}Bonnefoy, M., et al., 2011, A\&A, 528, 15L
%\bibitem[Boss(1997)]{Boss1997}Boss, A., 1997, Science, 276, 1836
%\bibitem[Bowler et al.(2010)]{Bowler2010}Bowler, B., et al., 2010, \apj, 723, 850
%\bibitem[Bromley and Kenyon(2011)]{BromleyKenyon2011}Bromley, B., Kenyon, S. J., 2011, \apj\ submitted
%\bibitem[Burgasser et al.(2002)]{Burgasser2002}Burgasser, A., et al., 2002, \apj, 571, 151L
%\bibitem[Burrows et al.(1997)]{Burrows1997}Burrows, A., et al., 1997, \apj, 491, 856
\bibitem[Burrows et al.(2006)]{Burrows2006}Burrows, A., et al., 2006, \apj, 640, 1063
%\bibitem[Chambers et al.(2010)]{Chambers2010}Chambers, J., O'Brien, D., Davis, A., 2010, in Protoplanetary Dust: Astrophysical 
%and Cosmochemical Perspectives, eds.: D. Apai, D. S. Lauretta, 2010, Cambridge University Press
%\bibitem[Chauvin et al.(2004)]{Chauvin2004}Chauvin, G., et al., 2004, A\&A, 425, 29L
%\bibitem[Chauvin et al.(2005)]{Chauvin2005}Chauvin, G., et al., 2005, A\&A, 438, 29L
\bibitem[Chen et al. (2008)]{Chen2008}Chen, C., Fitzgerald, M., Smith, P., 2008, \apj, 689, 539
\bibitem[Crepp and Johnson(2011)]{Crepp2011}Crepp, J., Johnson, J., 2011, \apj, 733, 126
\bibitem[Crifo et al.(1997)]{Crifo1997}Crifo, F., et al., 1997, A\&A, 320, 29L
%\bibitem[Currie et al.(2008)]{Currie2008}Currie, T., et al., 2008, \apj, 688, 597
%\bibitem[Currie(2009)]{Currie2009}Currie, T., 2009, \apj
\bibitem[Currie et al.(2009)]{Currie2009}Currie, T., Lada, C. J, et al., 2009, \apj, 698, 1
%\bibitem[Currie et al.(2010)]{Currie2010}Currie, T., Rieke, G. H., Kenyon, S. J., et al., 2010, \apj, in prep.
\bibitem[Currie et al.(2010)]{Currie2010a}Currie, T., et al., \apj, 721, 177L
\bibitem[Currie et al.(2011)]{Currie2011}Currie, T., Burrows, A., et al., 2011, \apj, 729, 128
%\bibitem[Deirmendjian(1964)]{Deirmendjian1964} Deirmendjian, D., 1964, Appl. Optics, 3, 187
%\bibitem[Duncan et al.(1998)]{Duncan1998}Duncan, M., et al., 1998, \aj, 116, 2067
\bibitem[Ehrenreich et al.(2010)]{Ehrenreich2010}Ehrenrieich, D., et al., 2010, A\&A, 523, 73 
%\bibitem[Fabrycky and Murray-Clay(2010)]{Fabrycky2009}Fabrycky, D., Murray-Clay, R., 2010, 710, 1408
\bibitem[Fitzgerald et al.(2009)]{Fitzgerald2009}Fitzgerald, M., et al., 2009, \apj, 706, 41L
%\bibitem[Ford and Chiang(2007)]{FordChiang2007}Ford, E., Chiang, E., 2007, \apj, 661, 602 
%\bibitem[Fukagawa et al.(2009)]{Fukagawa2009}Fukagawa, M., et al., 2009, \apj, 696, 1L
%\bibitem[Goldreich et al.(2004)]{Goldreich2004}Goldreich, P., Lithwick, Y., Sa'ari, R., 2004, ARAA, 42, 549
\bibitem[Golimowski et al.(2006)]{Golimowski2006}Golimowski, D., et al., 2006, \aj, 131, 3109
%\bibitem[Hillenbrand et al.(2002)]{Hillenbrand2002}Hillenbrand, L., et al., 2002, \pasp
\bibitem[Heap et al.(2000)]{Heap2000}Heap, S., et al., 2000, \apj, 539, 435
\bibitem[Hinkley et al.(2011)]{Hinkley2011}Hinkley, S., et al., 2011, \pasp, 123, 74
\bibitem[Hinz et al.(2006)]{Hinz2006}Hinz, P., et al., 2006, \apj, 653, 1486
%\bibitem[Hinz et al.(2010)]{Hinz2010}Hinz, P., et al., 2010, \apj, 716, 417
%\bibitem[Howard et al.(2010)]{Howard2010}Howard, A., et al., 2010, Science, 330, 653
%\bibitem[Hubeny and Burrows(2007)]{HubenyBurrows2007}Hubeny, I., Burrows, A., 2007, \apj, 669, 1248
%\bibitem[Huelamo et al.(2011)]{Huelamo2011}Huelamo, N., et al., 2011, A\&A, 528, 7L
%\bibitem[Ireland et al.(2010)]{Ireland2010}Ireland, M., et al., 2010, \apj\ in press, arXiv:1011.2201
\bibitem[Ireland and Kraus(2008)]{Ireland2008}Ireland, M., Kraus, A., 2008, \apj, 678, 59L
%\bibitem[Janson et al.(2010)]{Janson2010}Janson, M., et al., 2010, \apj, 710, 35L
\bibitem[Janson et al.(2011)]{Janson2011}Janson, M., et al., 2011, \apj, 728, 85
\bibitem[Kalas and Jewitt(1995)]{KalasJewitt1995}Kalas, P., Jewitt, D., 1995, \aj, 110, 794
\bibitem[Kalas et al.(2008)]{Kalas2008}Kalas, P., et al., 2008, Science, 322, 1345
%\bibitem[Kasper et al.(2007)]{Kasper2007}Kasper, M., Apai, D., Janson, M., Brander, W., 2007, A\&A, 472, 321
%\bibitem[Kenyon and Bromley(2009)]{KenyonBromley2009}Kenyon, S. J., Bromley, B., 2009, \apj, 690, 140L
%\bibitem[Kenyon and Bromley(2010)]{KenyonBromley2010}Kenyon, S. J., Bromley, B., 2010, \apjs, 188, 242
%\bibitem[King et al.(2010)]{King2010}King, \mnras
%\bibitem[Kratter et al.(2010)]{Kratter2010}Kratter, K., et al., 2010, \apj, 710, 1375
\bibitem[Kuchner and Stark(2010)]{Kuchner2010}Kuchner, M., Stark, C., 2010, \aj, 140, 1007
\bibitem[Lafreniere et al.(2007)]{Lafreniere2007a}Lafreniere, D., et al., 2007, \apj, 660, 770
%\bibitem[Lafreniere et al.(2007b)]{Lafreniere2007b}Lafreniere, D., et al., 2007b, \apj, 670, 1367
%\bibitem[Lafreniere et al.(2008)]{Lafreniere2008a}Lafreniere, D., et al., 2008, \apj, 689, 153L
%\bibitem[Lafreniere et al.(2008b)]{Lafreniere2008b}Lafreniere, D., et al., 2008b, \apj
%\bibitem[Lafreniere et al.(2009)]{Lafreniere2009}Lafreniere, D., et al., 2009, \apj, 694, 148L
%\bibitem[Lafreniere et al.(2010)]{Lafreniere2010}Lafreniere, D., et al., 2010, \apj, 719, 497
\bibitem[Lagrange et al.(2009a)]{Lagrange2009}Lagrange, A.-M., et al., 2009a, A\&A, 493, 21L
\bibitem[Lagrange et al.(2009b)]{Lagrange2009b}Lagrange, A.-M., et al., 2009b, A\&A, 506, 927 
\bibitem[Lagrange et al.(2010)]{Lagrange2010}Lagrange, A.-M., et al., 2010, Science, 329, 57
%\bibitem[Leggett et al.(2010)]{Leggett2010}Leggett, S., et al., 2010, \apj, 710, 1627
%\bibitem[Liu et al.(2010)]{Liu2010}Liu, M., et al., 2010, SPIE, 7736, 53
%\bibitem[Lunine et al.(1989)]{Lunine1989}Lunine, J., et al., 1989, \apj, 338, 314
\bibitem[Macintosh et al.(2008)]{MacIntosh2008}MacIntosh, B., et al., 2008, SPIE, 7015, 31
\bibitem[Madhusudhan et al.(2011)]{Madhusudhan2011}Madhusudhan, N., Burrows, A., Currie, T., 2011, \apj\ in press, arXiv:1102.5089
%\bibitem[Mamajek and Meyer(2007)]{Mamajek2007}Mamajek, E., Meyer, M., 2007, \apj, 668, 175L
%\bibitem[Marley et al.(2010)]{Marley2010}Marley, M. S., et al., 2010, \apj, 723, 117L
\bibitem[Marois et al.(2006)]{Marois2006}Marois, C., et al., 2006, \apj, 641, 556
%\bibitem[Marois et al.(2007)]{Marois2007}Marois, C., et al., 2007, \apj, 654, 151L
\bibitem[Marois et al.(2008)]{Marois2008}Marois, C., et al., 2008, Science, 322, 1348
\bibitem[Marois et al.(2010)]{Marois2011}Marois, C., et al., 2010, Nature, 468, 1080
\bibitem[Martinache and Guyon(2009)]{Martinache2009}Martinache, F., Guyon, O., 2009, SPIE, 7440, 20
%\bibitem[Masciardi et al.(2005)]{Masciardi2005}Masciardi, E., et al., 2005, \apj, 625, 1004 
%\bibitem[Metchev and Hillenbrand(2006)]{Metchev2006}Metchev, S., Hillenbrand, L. A., 2006, \apj, 651, 1166
%\bibitem[Metchev et al.(2009)]{Metchev2009}Metchev, S., Marois, C., Zuckerman, B., 2009, \apj, 705, 204L
\bibitem[Minowa et al.(2010)]{Minowa2010}Minowa, Y., et al., 2010, SPIE, 7736, 122
%\bibitem[Mizuno(1980)]{Mizuno1980}Mizuno, H., Prog. Theor. Phys., 64, 544
%\bibitem[Mohanty et al.(2007)]{Mohanty2007}Mohanty, S., et al., 2007, \apj, 657, 1064
%\bibitem[Moro-Martin et al.(2010)]{MoroMartin2010}Moro-Martin, A., Su., K., Rieke, G. H., 2010, \apj, 721, 199L
\bibitem[Mouillet et al.(1997)]{Mouillet1997}Mouillet, D., et al., 1997, \mnras, 292, 896
%\bibitem[Nielsen et al.(2010)]{Nielsen2010}Nielsen, E., et al., 2010, \apj, 
%\bibitem[Patience et al.(2010)]{Patience2010}Patience, J., et al., 2010, A\&A, 517, 76
%\bibitem[Pollack et al.(1996)]{Pollack1996}Pollack, J., et al., 1996, Icarus, 124, 62
\bibitem[Quanz et al.(2010)]{Quanz2010}Quanz, S., et al., 2010, \apj, 722, 49L
%\bibitem[Press et al.(1992)]{Press1992}Press, W., et al., 1992, \textit{Numerical Recipes in FORTRAN. 
%The art of scientific computing}, (Cambridge: University Press, --c1992, 2nd ed.)
%\bibitem[Rafikov(2004)]{Rafikov2004}Rafikov, R., 2004, \aj, 128, 1348
%\bibitem[Rafikov(2005)]{Rafikov2005}Rafikov, R., 2005, \apj, 621, 69L
%\bibitem[Rafikov(2010)]{Rafikov2010}Rafikov, R., 2010, \apj\ in press, arXiv:1004.5139
%\bibitem[Robitaille et al.(2007)]{Robitaille2007}Robitaille, T., et al., 2007, \apj, 129, 328
%\bibitem[Scholz et al.(2003)]{Scholz2003}Scholz, A., et al., 2003, A\&A, 
%\bibitem[Saumon et al.(2006)]{Saumon2006}Saumon, D., et al., 2006, \apj, 647, 552
\bibitem[Schneider et al. (2006)]{Schneider2006}Schneider, G., et al., 2006, \apj, 650, 414
\bibitem[Serabyn et al.(2010)]{Serabyn2010}Serabyn, E., et al., 2010, Nature, 464, 1018
\bibitem[Smith and Terrile(1984)]{SmithTerrile1984}Smith, B., Terrile, R. J., 1984, Science, 226, 1421
\bibitem[Thalmann et al.(2009)]{Thalmann2009}Thalmann, C., et al., 2009, \apj, 707, 123L
%\bibitem[Thalmann et al.(2010)]{Thalmann2010}Thalmann, C., et al., 2010, \apj
\bibitem[Weinberger et al.(2003)]{Weinberger2003}Weinberger, A., Becklin, E. E., Zuckerman, B., 2003, \apj, 584, 33L
\bibitem[Wilner et al.(2011)]{Wilner2011}Wilner, D., Andrews, S., Hughes, A. M., 2011, \apj, 727, 42L
\bibitem[Zuckerman et al.(2001)]{Zuckerman2001}Zuckerman, B., et al., 2001, \apj, 562, 87L
\end{thebibliography}
\end{document}